**Layered spin-crossover metal-organic frameworks for light-induced control of two-dimensional quantum materials**

*Carla Boix-Constant, Alejandro Orellana-Silla, José Antonio Real, Samuel Mañas-Valero,* Eugenio Coronado**

C. Boix-Constant, A. Orellana-Silla, J. A. Real, S. Mañas-Valero, E. Coronado
Instituto de Ciencia Molecular (ICMol), Universitat De València, Paterna, Spain

E-mail: samuel.manas@uv.es, eugenio.coronado@uv.es

Keywords: metal-organic frameworks, spin-crossover, 2D materials, van der Waals heterostructures, electronic transport, hybrid devices, quantum materials

## Abstract text

Metal-organic frameworks (MOFs) are widely explored for gas separation, catalysis or energy storage, yet their integration with other electronically functional materials remains largely unexplored. Here, we integrate the layered Hofmann-type spin-crossover (SCO) MOF $\{Fe^{II}(pyS_2Et)_2[Pt^{II}(CN)_4]\}$ into electrical nanodevices to control the properties of two-dimensional (2D) quantum materials. Our molecular approach exploits the stimuli-responsive nature of SCO layered materials as switchable building blocks. We demonstrate selective modulation of the electronic transport in van der Waals heterostructures interfacing SCO with 2D quantum materials (few-layer graphene, magnetic CrSBr, superconducting $NbSe_2$) through the strain induced *via* thermal and light-induced spin transitions. In graphene, the conductivity is selectively switched by light. In spin-valves based on CrSBr bilayers, the MOF triggers magnetic hysteresis (absent in pristine CrSBr) and enables tunable non-volatile zero-field memory. In $NbSe_2$, light modulates the superconducting critical current and transition temperature. These results establish layered stimuli-responsive SCO MOFs as active molecular control elements for 2D quantum materials, providing a route to strain-mediated control of electronic, magnetic, and superconducting functionalities and extending MOF-based architectures beyond traditional porous-matter applications towards multifunctional electronics and spintronics.

## 1. Introduction

Metal-organic frameworks (MOFs) are highly tunable materials whose chemical design allows us to tailor their properties for specific applications.[1–3] Examples include gas storage and separation, catalysis, chemical sensing and drug delivery.[4,5] All these applications exploit the porosity of these coordination polymers. However, their interest as quantum materials or their integration in electronics remains largely underdeveloped.[6–9] Recent efforts on electrically conductive MOFs[10,11] have demonstrated their versatility, including tunable conductivity *via* guest molecules, superconducting properties and proton conductivity.[12] Beyond their use as intrinsically conductive materials, an important opportunity offered by MOFs for electronics is their use as building blocks to induce desired modifications in electronic devices via interfacing with two-dimensional (2D) crystals forming van der Waals heterostructures.[13–15] For this purpose, the MOF must fulfil two requirements: (i) it must be mechanically exfoliable and stable enough to be mechanically transferred, and (ii) it must exhibit bistability that enables tuning of the 2D material's properties. As an example of exfoliable and bistable MOFs, we select a layered MOF exhibiting spin-crossover (SCO) transition, a paradigmatic case of molecular switching response.[16]

Pseudoctahedral iron(II) SCO complexes present two spin states – high-spin (HS, S = 2) and low-spin (LS, S = 0) – associated with the electronic configurations $t_{2g}^4e_g^2$ and $t_{2g}^6e_g^0$, respectively (**Figure 1a,b**), that can switch from one to another in presence of an external stimuli such as temperature, light, pressure, guest molecules, etc.[17,18] Given the antibonding nature of the $e_g$ orbitals, their population/depopulation during the spin-state switch is accompanied by a change of volume in the coordination core (usually $[FeN_6]$), being *ca.* 26% smaller for the LS state as it is the case of the paradigmatic SCO Hofmann-type MOF $[Fe^{II}(pyrazine)(Pt(CN)_4]$.[19,20] In the solid state, when the elastic coupling between the SCO centers is strong, a cooperative HS ↔ LS spin transition with thermal hysteresis can be observed (**Figure 1c**), conferring these magnetic materials their characteristic bistability.[17] When the SCO is interfaced with a 2D material, this volume change produces an effective local strain.[21] The SCO material can be combined with 2D materials by different approaches, such as the sublimation of the target molecules,[13,14] the deposition of nanoparticles mediated by colloidal suspensions,[21] drop-casting,[22] micro-contact printing,[23] or *via* the assembly of van der Waals heterostructures (vdWHs) based on layered SCO MOFs, which can be mechanically exfoliated even down to the single-layer limit.[24] A summary of different layered SCO MOFs interfaced with 2D materials is given in the **Supplementary Information Table S9**.

Key advantages of the molecular nature of layered $Fe^{II}$ SCO MOFs are their chemical stability and design versatility. Thus, SCO parameters such as transition temperature, thermal hysteresis wideness, number of switching steps, etc., can be chemically tuned through isomorphic substitution of the SCO metallic centers, by changing the nature of coordinating ligands, or even by the inclusion of appropriate guest molecules in the structure.[25] Furthermore, the SCO behaviour can also be physically perturbed by the application of pressure and even by light irradiation at low temperatures.[26] In particular, the light-induced switching between spin states is known as the LIESST effect (LS→HS*; LIESST = Light-Induced Excited Spin State Trapping), where HS* is a metastable photo-generated state that can persist for long periods of time at low temperatures (**Figure 1b, c**).[27,28] This process is accompanied by the reverse-LIESST effect (HS*→LS)[29] upon warming, when the system relaxes back to the LS state at a characteristic temperature known as $T_{LIESST}$. The value of $T_{LIESST}$ depends on the system and enables reversible switching through alternating light irradiation and thermal relaxation (**Figure 1b,c**).[29,30] The LIESST effect has been employed for switching the graphene resistivity when incorporated on hybrid molecular/inorganic heterostructures,[13,31] opening the door to the control of other quantum materials (as magnets or superconductors). Recently, its unmatched potential for controlling spintronic and magnonic devices has been theoretically proposed,[32] although its observation still remains experimentally unaddressed.

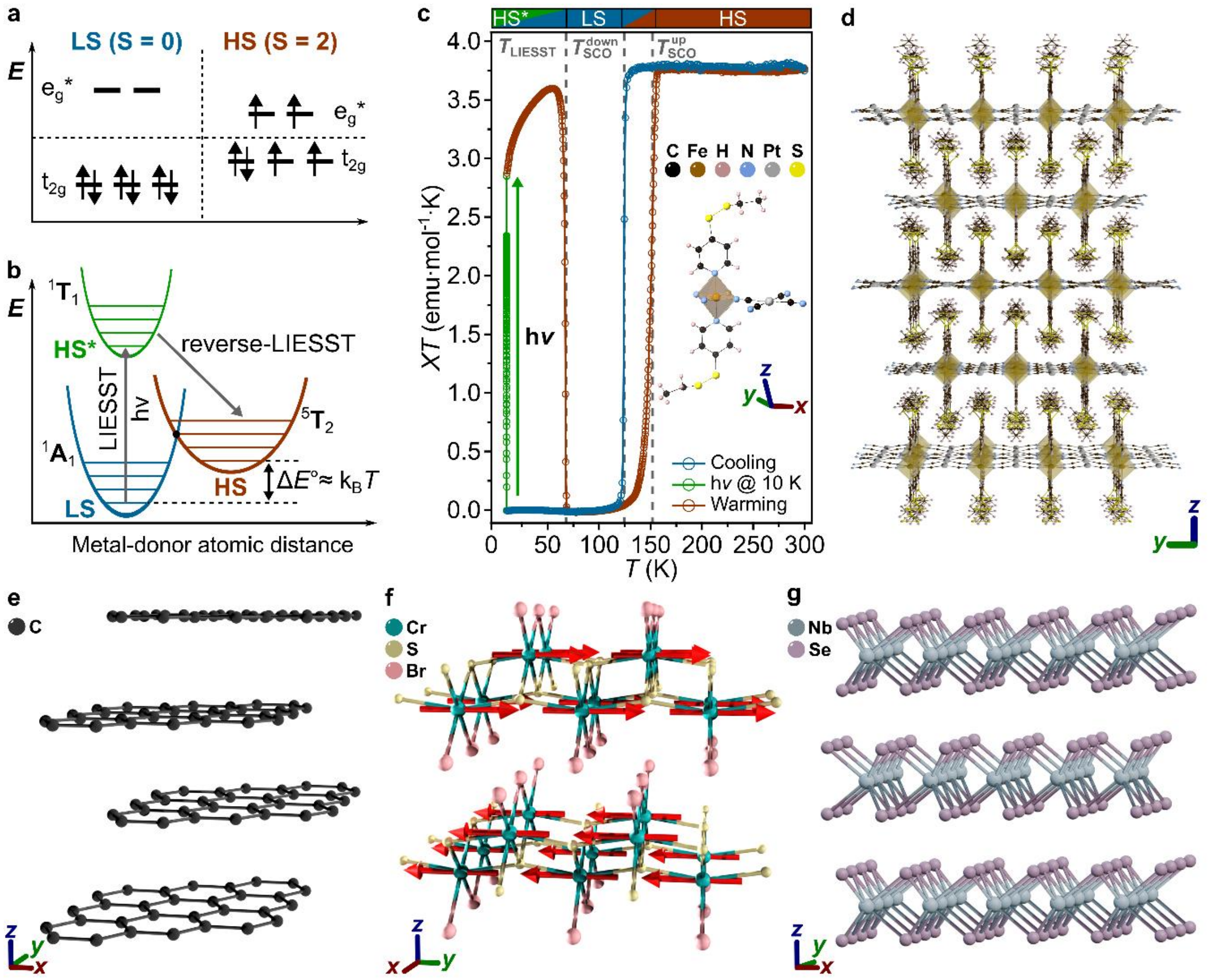

**Figure 1. Layered spin-crossover (SCO) metal-organic framework (MOFs). a**, Electronic configurations of the low-spin (LS) and high-spin (HS) states of the $Fe^{II}$ metal centres in an octahedral ligand field, showing the occupation of the HOMO and LUMO orbitals. The SCO transition involves a reversible switch from a paramagnetic HS state (S = 2) at high temperatures to a diamagnetic LS state (S = 0) at low temperatures, accompanied by significant changes in the lattice parameters (see **Table S8** in the **Supporting Information Section 3**). **b**, Jablonski diagram illustrating the SCO phenomenon, depicting the electronic transitions and the light-induced excited spin state trapping (LIESST) effect. **c**, Magnetic and photomagnetic properties of $\{Fe^{II}(pyS_2Et)_2[Pt^{II}(CN)_4]\}$ (*FePtpyS₂Et*). The blue, red and green curves correspond to the cooling, heating and photo-switching process (light irradiation at 10 K), respectively. The color bar at the top illustrates the SCO electronic states at different temperatures: red for HS, blue for LS, and green for HS*. The temperature regions where two spin states coexist (bistable SCO regime) are represented by a combination of the corresponding colors. Inset: detail of the $Fe^{II}$ SCO centres. **d**, Side view of the crystal structure of *$FePtpyS_2Et$*. **e-g**, Structures of the inorganic 2D materials used in this study: (e) few-layers graphene; (f)

vdW magnet CrSBr (red arrows represent the spins along the easy magnetic *b*-axis); (g) superconducting transition metal dichalcogenide $NbSe_2$.

## 2. Results and Discussion

### 2.1. Layered MOFs with light-induced SCO transition for photo-responsive vdWHs

*FePtpyS₂Et* belongs to an extensive series of 2D Hofmann clathrates where the $Fe^{II}$ nodes are equatorially surrounded by four square-planar $[M^{II}(CN)_4]^{2-}$ bridging building blocks ($M^{II}$ = Ni, Pd, Pt), thereby defining $\{Fe_2[M^{II}(CN)_4]_2\}_\infty$ grids, while the axial positions of the $Fe^{II}$ octahedron are filled by two unidentate terminal ligands, *pyS₂Et* in the present case.[17,18,33] This material presents a strong cooperative SCO transition featuring a thermal hysteresis 32.5 K wide with characteristic SCO temperatures of $T_{SCO}^{down}$ = 121.5 K and $T_{SCO}^{up}$ = 154 K for the cooling and heating modes, respectively (**Figure 1c**). Furthermore, upon light irradiation at low temperatures ($\lambda$ = 532 nm, typical for LIESST switching in $Fe^{II}$ SCO compounds; see **Experimental Section**), the compound quantitatively switches from the LS state to the metastable HS* state, which subsequently relaxes back to the stable LS state at $T_{LIESST} \approx 69$ K upon heating in the dark (**Figure 1c**). A key requirement for incorporating the magnetic phenomenology described above into vdW devices is the isolation and manipulation of thin MOF layers. Thanks to the layered crystal structure of the MOF (**Figure 1d**), micromechanical exfoliation can be effectively used to obtain thin-layers, which can be deterministically manipulated using state-of-the-art techniques developed for 2D materials. All these characteristics enable the fabrication of photo-responsive vdWHs based on layered SCO MOFs.[34,35]

The volume change accompanying the SCO transition offers a general strain-engineering handle that is independent of the specific electronic order of the interfaced 2D material. To establish this versatility, we systematically probe three distinct electronic responses (charge transport, magnetism, and superconductivity) using few-layer graphene (FLG; **Figure 1e**), CrSBr (**Figure 1f**) and $NbSe_2$ (**Figure 1g**), respectively, as prototypical and complementary 2D quantum materials. As detailed below, this progression allows us to benchmark the strain effect triggered at low temperatures by a light-induced spin transition in a comparatively simple electronic system (**Section 2.2**), before extending it to strongly correlated systems as 2D magnets (**Section 2.3**) and superconductors (**Section 2.4**). Here we have chosen a SCO material exhibiting LIESST effect, since, in contrast to what happens with a thermally-induced spin transition, this light-induced spin transition occurs at temperatures that are compatible with the electronic order of these quantum materials.

## 2.2. Selective electrical resistivity switch in SCO/FLG vdWHs

The electrical properties of graphene are highly sensitive to the strain-induced by the SCO transition.[13,15,23] We benchmark the thermal and light-induced spin transition of mechanically exfoliated *FePtpyS$_2$Et* by using FLG as the sensing 2D material. The SCO/FLG vdWHs consist of a SCO flake placed between gold electrodes that is covered with a top FLG (**Figure 2a,b**). By measuring the electrical resistivity, we sense the thermally induced SCO transition at the expected transition temperatures of the clathrate, which manifests as a clear hysteresis in the heterostructure (**Figure 2c**, top panel), being absent for the bare FLG (**Figure 2d**, top panel).

In addition to the thermal spin-transition, upon light irradiation at 2 K we observe a resistivity decrease, corresponding to the transition from LS – more resistive (blue trace in **Figure 2c,d**, bottom panel) – to HS* – less resistive (red trace in **Figure 2c,d**, bottom panel). When warming up, at $T_{\mathrm{LIESST}} \approx 69$ K, an increase of the resistivity is observed, pinpointing the transition from HS* back to LS. This temperature is comparable to the bulk one (**Figure 1c**), demonstrating the persistence of the light-induced SCO phenomena in the thinner crystals. Furthermore, we showcase the selectively light-induced resistance switch by irradiating at different temperatures below $T_{\mathrm{LIESST}}$ to reach HS* (see **Experimental Section** for technical details, and **Supporting Information Section 2.1.3.** for different exposure times and laser powers), observing a clear resistance drop after light irradiation and a systematic resistance enhancement at $T_{\mathrm{LIESST}}$ while warming up (**Figure 2e**). Several phenomena could in principle contribute to the observed effect, including strain effects, charge transfer, contact resistance changes, photothermal effects, and laser-induced changes. Our measurement scheme, based on a four-probe configuration with simultaneous measurement of the vdWH and the bare FLG as a reference, in which no hysteretic response is observed, allows us to rule out contact resistance changes, photothermal effects, and laser-induced changes as the dominant origin. Regarding charge transfer, this mechanism is expected in SCO materials when hybridization at the molecular/2D interface occurs. This is the case, for example, for SCO molecules deposited by sublimation or evaporation methods on metallic surfaces. [36] In our case, however, our transfer method (mechanical stacking of one layer onto another), together with the insulating nature of the SCO and the good electronic insulation provided by the ligands surrounding the $Fe^{II}$ metal ion, make this hybridization very unlikely, as previously shown. [15]

To further experimentally discard this origin, we decouple now the two systems by introducing a thin insulating hBN layer between the SCO and the FLG, fabricating a SCO/hBN/FLG heterostructure (device shown in **Figure S5;** see **Supplementary Information Section 2.1.4** for the full electric characterization of the decoupled device). Even in this decoupled heterostructure, we still observe both the thermal- and light-induced transitions, supporting a mechanical strain origin for the effect rather than a charge-transfer mechanism. Notably, it is also important to mention that the effect is reproducibly observed across devices with different relative lattice orientations between the SCO MOF and the 2D material, consistent with a strain mechanism driven by the SCO volume change rather than by lattice-matching effects.

We observe consistent results across different devices, shown in **Supporting Information Section 2.1** (thicknesses and geometrical factors are summarized in **Table S1**). As previously reported with similar systems,[15] we perform several thermal cycles as well as I-V curves to check the robustness of the effect (up to 20 thermal cycles for the SCO/FLG devices). Still, we note a higher fatigue in $FePtpyS_2Et$ if compared with the related Hofmann clathrate {Fe-(3-Clpyridine)$_2$-[$Pt^{II}(CN)_4$]},[15] observing several cracks on the crystal upon cooling (see **Figures S6** and **S8**), thus motivating further chemical efforts for designing robust layered SCO MOFs. Having established that SCO-induced strain reliably and reversibly modulates charge transport in a benchmark 2D system as FLG, we next examine whether the same mechanism can control a qualitatively different electronic property: long-range magnetic order.

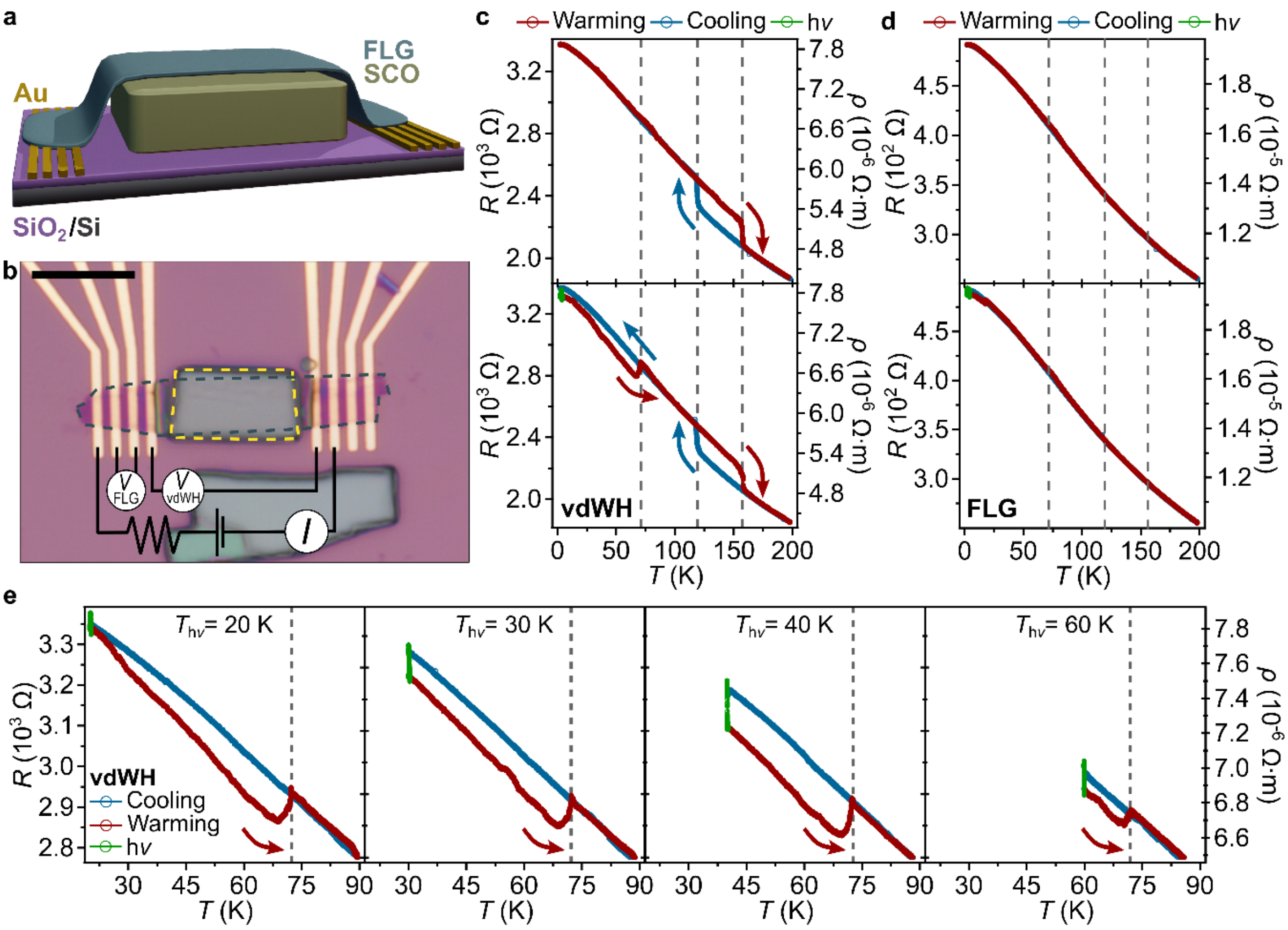

**Figure 2. Selective light-induced electrical resistivity switch in SCO/FLG vdWHs. a**, Sketch (not to scale) of the vdWH for SCO/FLG devices, where a FLG flake is placed on top of the SCO crystal. **b**, Optical image of a representative device (*vdWH A.1.* in **Supporting Information Section 1.2.1**) of FLG (gray dashed lines) placed on top of a SCO MOF (yellow dashed lines), with the electrical configuration sketched. Scale bar: 10 µm. **c,d**, Thermal spin transition (top panels) and light-induced switch at 2 K (bottom panels) for the vdWHs (**c**) and the bare FLG (**d**). **e**, Thermal dependence of resistance/resistivity when irradiating at selected temperatures (indicated as $T_{hv}$ in the panels). A consistent resistance switch with a diminishment is observed after light irradiation, followed by an enhancement at $T_{LIESST} \approx 69$ K while warming up. Electrical characterization of different devices, light-irradiation times, temperatures and powers are shown in **Supporting Information Section 2.1**.

### 2.3. Emergence of magnetic hysteresis in SCO/bilayer CrSBr vdWHs

The switching behaviour shown before can be employed for tuning other 2D materials; for instance, 2D magnets. We demonstrate the effect in CrSBr, a semiconducting vdW metamagnet, exhibiting A-type antiferromagnetic ordering at $T_N = 132$ K.[37] This material has recently gained a lot of interest due to the interplay between its electronic, optical and magnetic properties, which can be studied down to the single layer and tailored *via* twist, stack or strain engineering.[38–42] In bulk, the spins are ferromagnetically aligned within the layers, which couple antiferromagnetically between them. The field-induced magnetic transitions in CrSBr are very sensitive to external strain,[41,43,44] and depend on the magnitude and relative orientation of the applied magnetic field with respect to the magnetic axes, where the easy, intermediate and hard magnetic axes correspond to the crystallographic *b*, *a* and *c* directions, respectively.

In **Figure 3a** we show a typical device formed by a CrSBr bilayer (2L) placed on pre-lithographed gold contacts with a SCO layer on top. We studied the bilayer since it represents the thinnest case for a spin-valve-like response when an external magnetic field is applied, characterized by low and high resistance states when the layers' magnetizations are parallel and antiparallel, respectively.[37] The spin-reversal process at 10 K is strikingly different for the vdWH (**Figure 3b**). While the pristine bilayer exhibits a continuous spin reorientation until reaching saturation (**Figure 3c**), the vdWH presents a marked magnetic hysteresis, as better visualized by non-zero $\Delta R$ values (where $\Delta R = R_{+B\rightarrow-B} - R_{-B\rightarrow+B}$; top panels in **Figure 3b,c**). In addition, the saturation field to achieve the parallel configuration in the vdWHs (*ca.* ± 3 T) occurs at higher magnetic fields compared to the pristine bilayer (*ca.* ± 2 T).[37] The full field dependence curve up to ± 9 T is shown in **Figure S21**. This phenomenology aligns with the higher saturation fields observed when biaxial compressive strain is applied to thin CrSBr layers (~ 10 nm), predicted to arise from enhanced interlayer exchange coupling and increased magnetic anisotropy energy.[38] Still, in our case, the observed hysteresis may be attributed to the formation of magnetic domains, as we experimentally address by performing FORCs. We note that this origin does not permit a direct comparison with the previously developed theoretical models, thereby motivating both further theoretical investigations and magnetic imaging experiments to elucidate the interplay between strain and magnetic domain formation in SCO-based heterostructures.

To study the irreversibility of the magnetic process, we perform first-order reversal curves (FORCs).[39,40] To initialize the system, we first saturate at - 9 T and, then, we sweep the field to $B$ = - 1.5 T. From this value, we increase the field sequentially by steps of 20 mT ($B_{\max}$), sweeping back to $B$ = - 1.5 T after each $B_{\max}$ is reached (field protocol sequence: -1.5 T → $B_{max}$ → -1.5 T). Selected $B_{\max}$ values are shown in **Figure 3d**, whereas the full dataset is presented in **Figure 3e**. At low $B_{\max}$ values, the forward (-1.5 T → $B_{max}$; red trace) and backward ($B_{max}$ → -1.5 T; blue trace) curves follow the same pathway, presenting a reversible behaviour with no hysteresis ($\Delta R$ = 0, as shown in the top panel of **Figure 3d**). Above a certain field threshold ($B_{\max}$ = - 0.02 T for the device shown in **Figure 3**), the forward and backward traces show an irreversible behaviour, with the appearance of magnetic hysteresis ($\Delta R \neq 0$). Notably, non-volatile memory at zero-field is observed by the proper field protocol, as shown by selecting $B_{max}$ = 0.5 T in **Figure 3d**. Additionally, the resistance exhibits sharp changes at certain magnetic fields (**Figure 3e**). This phenomenology can be ascribed to the switching of magnetic domains, as observed in orthogonally-twisted CrSBr layers.[39,40,45] Finally, we consider the LIESST effect by comparing the hysteresis in the FORC measurements before light irradiation – LS state – and after light irradiation – HS* state –, observing that the intensity of the effect ($\Delta R$) can be modulated (although weakly) by light irradiation, with differences more significant for certain applied $B_{max}$ fields (**Figure 3f**; raw resistance data shown in **Supporting Information Section 2.2.1**). The weak light dependence suggests that the first thermal spin transition configures in a non-reversible manner the magnetic domain landscape. Beyond charge transport and magnetism, we finally probe whether SCO-induced strain can also perturb a macroscopic quantum coherent phase as superconductivity, using $NbSe_2$ as a model system.

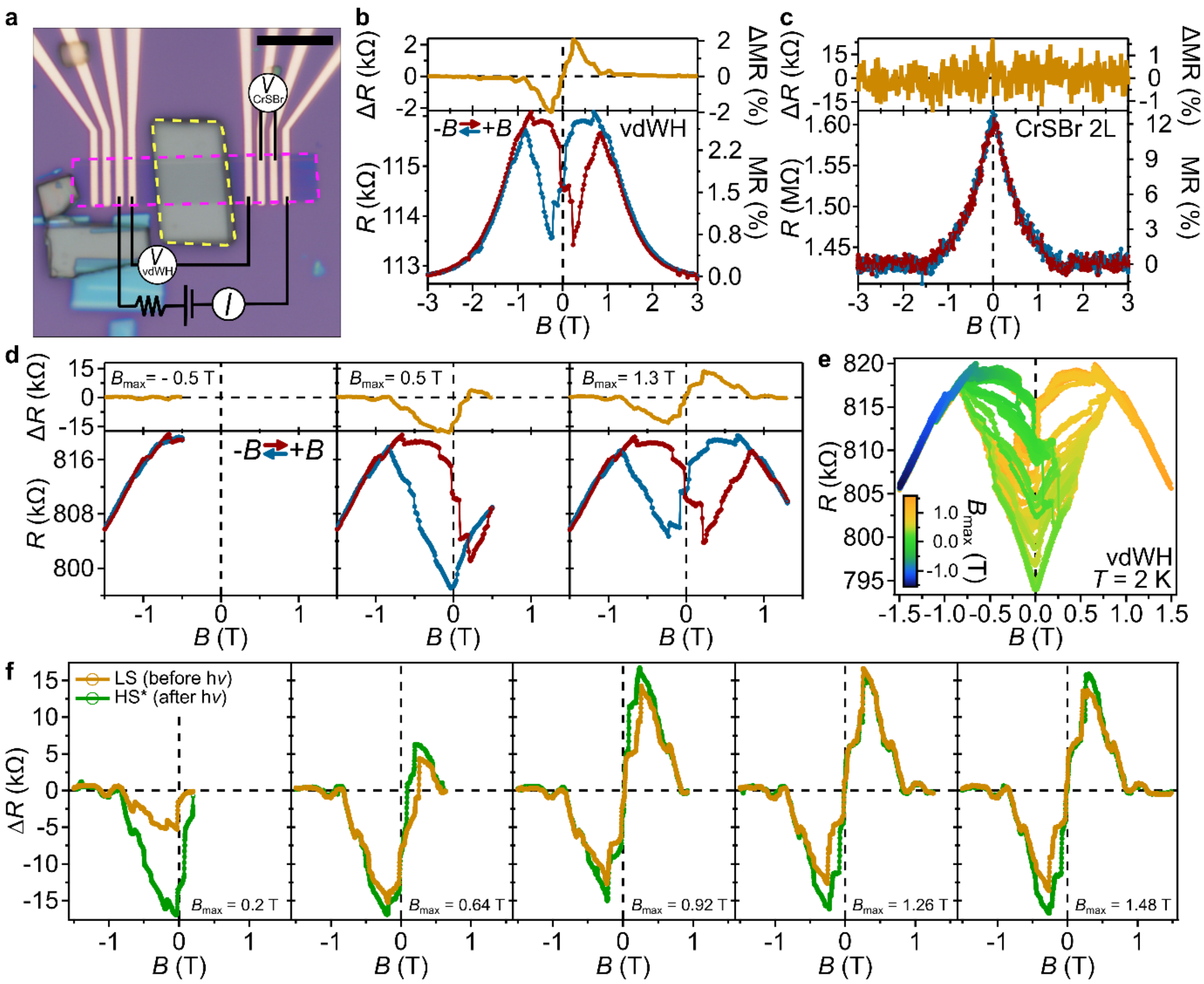

**Figure 3. Magnetic hysteresis opening in SCO/CrSBr (2L) vdWHs. a**, Optical image of a representative device (*vdWH B.1.* in **Supporting Information Section 2.2.1**) based on a SCO

MOF (yellow dashed lines) placed over a bilayer of CrSBr (pink dashed lines), with the corresponding electrical configuration sketched. Scale bar: 10 µm. **b,c**, Field-dependence of the resistance/magnetoresistance (bottom panel) with its difference (top panel) for both the heterostructure (**b**) and the reference pristine CrSBr bilayer (**c**) at 10 K, with applied fields in the out-of-plane direction ($B.| c$). $\Delta R$ is defined as $\Delta R = R_{+B\rightarrow -B} - R_{-B\rightarrow +B}$, and MR (%) = $100\cdot[R(B) - R(\mathrm{P})]/R(\mathrm{P})$, where $R(\mathrm{P})$ is the resistance of the layers in the parallel configuration (when saturation is reached). **d**, First-order reversal curves (FORCs) for selected $B_{\mathrm{max}}$ values at 2 K. **e**, Full set of FORCs for the vdWH at 2 K. **f**, Light-induced $\Delta R$ effects before (yellow trace) and after (green trace) light irradiation at 2 K with the field applied parallel to the *c* axis.

### 2.4. Superconducting critical parameters switch in SCO/$NbSe_2$ vdWHs

Having demonstrated SCO-induced control over charge transport and magnetism, we turn to a third and more stringent test case: a superconducting phase, where strain-induced perturbations, if present, would manifest as measurable shifts in well-defined critical parameters ($T_c$, $I_c$). Transition metal dichalcogenides (TMDs) exhibit a wide variety of electrical and optical behaviours, allowing their exploitation in different applications in fields as diverse as optoelectronics or energy storage, among others.[46] Strain effects are well-studied in this family of materials, in particular for the semiconducting systems, where strain-induced modifications in their band structure are observed.[42] However, beyond semiconducting TMDs, the number of strain-effects studies on metallic and superconducting TMDs are significantly lower. Here, we select the prototypical metallic TMD 2H-$NbSe_2$, which exhibits superconductivity at low temperatures with a critical temperature ($T_c$) of 7.2 K in bulk that decreases in thin-layers.[47]

We integrate $NbSe_2$ thin layers (thicknesses ≈ 20 nm; see **Supporting Information Section 1.2**) with the SCO MOF as shown in **Figure 4a**. In the metallic region, the thermal spin transition does not show a significant impact in the $NbSe_2$ resistivity (**Supporting Information Section 2.3**), as expected for a band metal and in agreement with previous reports.[15] Still, the spin transition has an impact on the superconducting properties, as revealed by triggering the LIESST effect. In particular, we observe in the vdWH subtle changes in the $T_c$ upon the light-induced spin-transition (**Figure 4b**), where we perform several thermal cycles between 2 K and 8 K before and after light-irradiation at 2 K (temperature sweeping rate of 0.5 K/min; see **Experimental Section** for further details). The $T_c$ change ($\Delta T$), where the $T_c$ is defined following the 10%, 50% and 90% criteria (**Experimental Section**), is almost null at $T_c^{10\%}$ ($\Delta T^{10\%}$ = 0.014 K), starts increasing at $T_c^{50\%}$ ($\Delta T^{50\%}$ = 0.043 K), and is further enhanced at $T_c^{90\%}$ ($\Delta T^{90\%}$ = 0.06 K). Despite being the overall $T_c$ change small, the effect is reproducible over different thermal cycles and devices (see **Supporting Information Section 3.3**, where we summarize the results for four different devices). Comparing these results with the bare $NbSe_2$ layer (**Figure 4c**), no change on the $T_c$ is observed upon light irradiation.

We consider now the modulation in the critical current ($I_c$), defined as the current value where the maximum of the differential resistance, *dV/dI,* is observed.[48] We show in **Figure 4d** the I-V curves as well as the d*V*/d*I* at 2 K. Upon light irradiation, the vdWH exhibits a critical current change enhancement ($I_c \approx$ 0.290 mA before light irradiation, and $I_c \approx$ 0.558 mA after light irradiation), as indicated by the grey dashed lines in **Figure 4d**. By tracking the d*V*/d*I* at different temperatures (**Figure 4e**), we observe that the light-irradiation – and, therefore, the underlying local strain due to the transition from LS to HS* – induces a noticeable modification in the critical current of the vdWH, while $T_c$ is only marginally modified (in line with the experimental observation in **Figure 4b**); as comparison, the pristine $NbSe_2$ does not show any light-induced modification (black dashed lines in **Figure 4d**). Taking the values of $I_c$ extracted from **Figure 4e**, we fit them (**Figure 4f**) following the model by Thompson *et al.*[49] where $I_c(T) = I_c(0)\cdot[1-(T/T_c)^2]^n$, being $I_c(0)$ the critical current at 0 K, $T_c$ the critical temperature, and $n$ an empirical fitting parameter determined by the vortex pinning mechanism (typical values range from $n$ = ½ to $n$ = 2, corresponding to a weak and strong pinning cases, respectively). For the pristine $NbSe_2$ flake, we obtain $n \approx$ ½, both before (0.504 ± 0.009) and after light irradiation (0.505 ± 0.008), indicative of weak pinning.[50] Considering the vdWH, there is an increase of $n$ after light irradiation from 0.575 ± 0.014 to 0.696 ± 0.013, with $\Delta n$ = (0.121 ± 0.019), suggesting a stronger vortex pinning scenario that may be ascribed to local modification in terms of ripples and strain induced by the spin transition, in line with previous observations in $NbSe_2$.[47,51,52] In particular, Wieteska *et al.*[53] observed a $T_c$ decrease upon uniaxial tensile strain. In our case, the SCO transition from LS to HS* implies a volume expansion, thus triggering also a tensile strain. Still, we note that the situation in our experiments may be more complex, as the SCO transition could induce different local strain levels across the heterostructure, potentially leading to local variations in vortex pinning. All the fitting parameters for the measured samples are summarized in the **Supporting Information Section 2.3**.

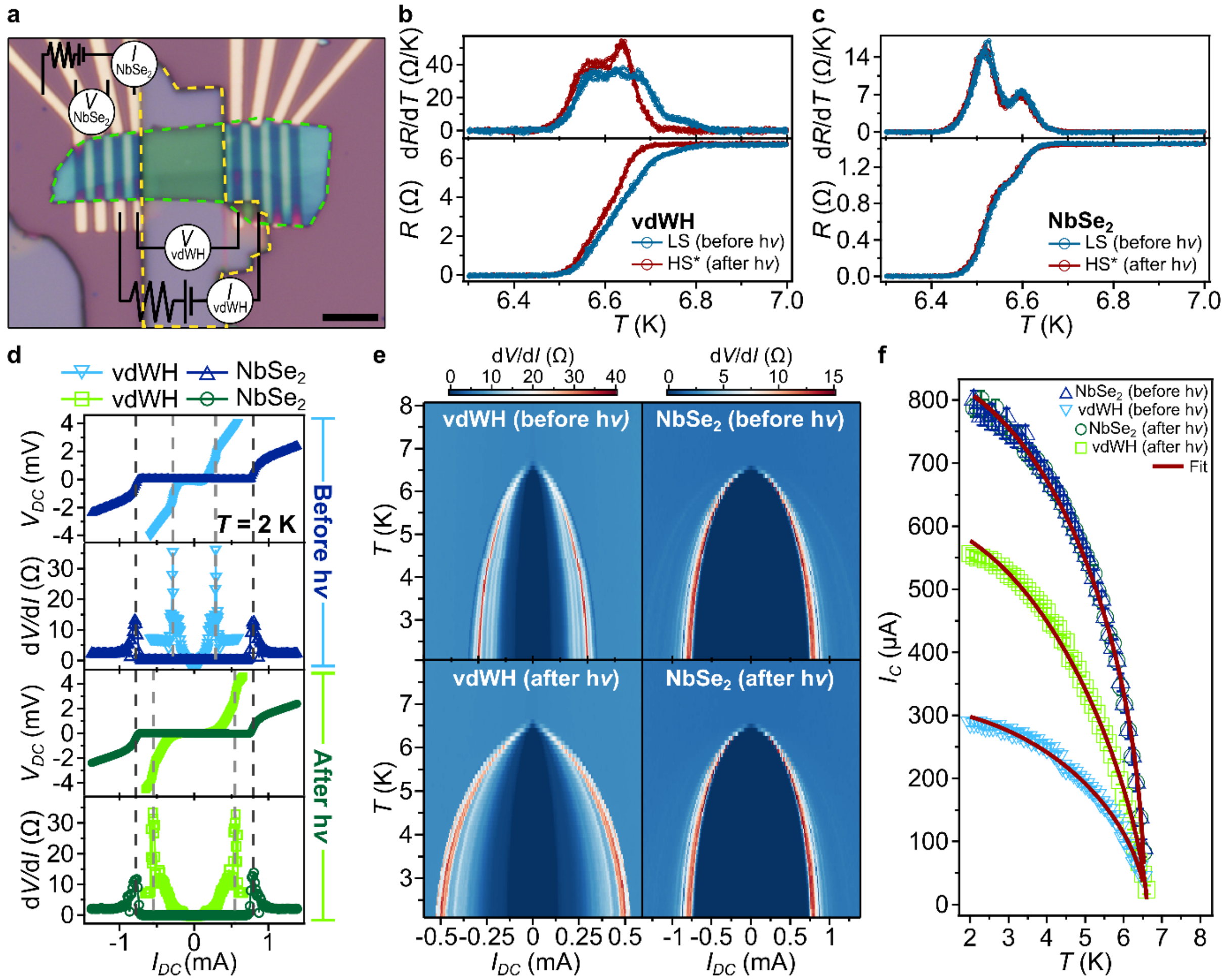

**Figure 4. Light-induced superconducting critical temperature and current modulation in SCO/$NbSe_2$ vdWHs. a**, Optical image of a representative device (*vdWH C.1.* in **Supporting Information Section 2.2.1**) based on a SCO MOF (yellow dashed lines) on top of a thin-layer of $NbSe_2$ (green dashed line), with the corresponding electrical configuration sketched. Scale bar: 10 μm. **b,c**, Temperature dependence of the resistance and its derivative, d$R$/d$T$, for the heterostructure (**b**) and reference $NbSe_2$ (**c**) before (blue trace) and after (red trace) light irradiation at 2 K. **d**, I-V curves and differential resistance at 2 K for both the vdWH and $NbSe_2$ before (blue traces) and after (green traces) light irradiation. **e**, Temperature dependence of the differential resistance, d$V$/d$I$, as a function of the DC current for the reference $NbSe_2$ (left panels) and the vdWH (right panels), before (top) and after (bottom) light irradiation. **f**, Evolution of the superconducting critical current as a function of temperature for both the vdWH and the reference $NbSe_2$ layer, before and after light irradiation. The fitting follows the thermal dependence described as $I_c(T) = I_c(0)\cdot[1-(T/T_c)^2]^n$, where $I_C(0)$ is the critical current at 0 K, $T_c$ is the critical temperature for each device, and $n$ is an empirical fitting parameter.[49] Electrical transport data corresponds to device *vdWH C.1.* (**Supporting Information Section 2.3.1**)

## 3. Conclusion

In this work, we establish that the strain generated by a light- and thermally-switchable SCO MOF constitutes a general strategy for controlling electronic properties in 2D quantum materials. We demonstrate this principle across three distinct classes of electronic order (charge transport, magnetism, and superconductivity) by deterministically integrating thin-layers of a SCO MOF exhibiting light-induced magnetic bistability with few-layer graphene, 2D magnets and superconductors, incorporating them into electrical nanodevices. In particular, our molecular strategy harnesses the intrinsic volume change concomitant to the spin transition for strain engineering. In few-layers graphene, we selectively switch between high/low resistivity states controlled upon light irradiation. In 2D magnets, we trigger the emergence of magnetic hysteresis in bilayer CrSBr – absent in the pristine CrSBr bilayer case – enabling a tuneable switch between zero-field non-volatile and volatile memory that is controlled *via* the applied field protocol and further modulated by light. In layered superconductors, we observe light-induced modulations of the superconducting critical temperature and critical current in $NbSe_2$ thin-layers. Considering the thickness dependence of the 2D material, we use few-layer graphene for its easier exfoliation, allowing us to benchmark the SCO MOF response; bilayer CrSBr, as the thinnest case where a spin-valve-like response is observed; and thin-layer $NbSe_2$, to study the superconducting transition above 2 K. The SCO-induced transition temperatures in FLG match those of bulk MOF regardless of FLG thickness. In addition, the thickness dependence of CrSBr and $NbSe_2$ could be exploited to tailor the device properties: thicker CrSBr flakes yield a higher coercive field (at 2 K, increasing from 0.2 T in the bilayer to 0.6 T in bulk samples, for fields along the easy magnetic axis),[37] while thinner $NbSe_2$ flakes would allow the superconducting transition temperature to be tuned (from 7.2 K in bulk down to *ca.* 1 K in the monolayer limit).[54]

More broadly, our results demonstrate the potential of incorporating molecular building blocks into van der Waals heterostructures to actively control quantum materials. Specifically, the selected layered SCO MOF provides access to physical phenomenology *via* the LIESST effect that enables optical control of otherwise light-insensitive 2D crystals. Thanks to the chemical tunability of MOFs, spin transition temperatures and lattice parameters can be designed, providing significant opportunities for device engineering with optimized performance.

Beyond SCO systems, we envision that this molecular strategy can be further extended to other photoactive mechanisms beyond the LIESST effect, including light-induced charge transfer[55] or reversible bidirectional photoswitching processes, [56] therefore opening chemical strategies for selectively tuning 2D materials and device functionalities. This molecular approach offers novel control strategies, providing a complementary route to conventional gating, stacking, and twisting techniques. Moreover, the vdWHs approach facilitates the integration of a wide variety of materials with diverse properties (including metals, superconductors, magnets, and topological insulators).

Overall, we highlight the versatility and potential of MOFs well beyond their conventional applications, demonstrating how their integration into hybrid molecular/2D heterostructures enables a platform for active control in nanoelectronics, spintronics or data storage.

## 4. Experimental Section

*Crystal growth:* $NbSe_2$ and CrSBr crystals were grown using chemical vapour transport (CVT) method,[37,57] whereas $\{Fe^{II}(pyS_2Et)_2[Pt^{II}(CN)_4]\}$ SCO material was obtained from slow diffusion techniques in water-methanol solutions as reported previously.[58] Graphite was purchased in NGS Naturgraphit GmbH.

*Bulk magnetic measurements:* Variable-temperature (10 - 300 K) direct current (DC) magnetic susceptibility measurements were carried out under an external field of 0.1 kOe, with a cooling/heating rate of 1 $K\cdot min^{-1}$. For the light-induced state, the bulk SCO crystals were exposed to light irradiation at 10 K for 220 min, with an effective laser power of 640 μW and a relaxation rate of 0.3 $K\cdot min^{-1}$. All the magnetic measurements were performed with a SQUID magnetometer (Quantum Design model MPMS3). LIESST timescale is shown in the **Supplementary Information Section 1.1**.

*Fabrication of vdW heterostructures:* Thin-layers of the van der Waals materials were obtained using the Scotch-tape method, and the assembly of the heterostructures was performed using a dry deterministic transfer approach employing poly-(bisphenol A carbonate)/polydimethylsiloxane (PC/PDMS) stamps. Electrical gold contacts were pre-patterned on top of high-resistive $SiO_2$ (300 nm)/Si substrates (from NOVA Electronic Materials) using standard electron-beam lithography techniques, followed by the evaporation of the metallic contacts (2 nm/75 nm Cr/Au) from a thermal source in a vacuum chamber (Angstrom Engineering Nexdep) inside a glovebox, with a base pressure of $7 \cdot 10^{-7}$ mbar. The electrical design allows us to perform 4-probe measurements both at the heterostructure area (labelled as *vdWH*) and at the reference 2D inorganic system (labelled as FLG, CrSBr (2L) and $NbSe_2$, respectively).

*Electrical characterization:* Electrical measurements were performed using a Physical Property Measurement System PPMS-DynaCool, from Quantum Design (2 - 400 K; $B = \pm 9$ T), with a 4-probe geometry where a DC current is applied through the outer leads, and the voltage drop is measured at the inner contacts. AC/DC voltages and currents were measured with a lock-in (MFLI from Zurich Instruments) using an external resistance of 10 MΩ, *i.e.*, a resistance much larger than that of the sample. Temperature sweeps were performed at 1 $K \cdot min^{-1}$ (FLG and CrSBr vdWHs) and 0.5 $K \cdot min^{-1}$ ($NbSe_2$ vdWHs), and the field sweeps (type B vdWHs) were performed at 200 $Oe \cdot s^{-1}$. For differential resistance measurements, the characteristic DC I-V curves were tracked at zero field as a function of the temperature applying simultaneously a small AC excitation ($V_{AC} = 50$ mV, $f = 27.7$ Hz). $I_C$ was extracted as the current value for the maximum dV/dI. For defining the critical temperature ($T_C$) of both the vdWHs and the $NbSe_2$ thin layers, we follow the 10%, 50% and 90% criteria – that is, the temperatures at which the electrical resistance dropped to 10%, 50% and 90% with respect to the normal-state resistance ($R_n$) for both the LS and HS* states, where $R_n$ was taken just above the superconducting transition for each device.

*Light irradiation:* Light-induced experiments were performed using a multimode fibre – core diameter of 400 μm, 0.50 NA, with a SMA connector on one end, and a 2.5 mm ceramic ferrule on the other end from Thorlabs (reference M126L01). The fibre goes along the hollow rod of a home-made cryostat inset, having a SMA feedthrough on the top, connected to a green laser source ($\lambda = 532$ nm, from Roithner Lasertechnik GmbH, model RLTMGL-532-100-3); the bottom ferrule is kept in place and aligned so the laser spot is over the sample in a normal incidence. Our standard light-irradiation protocol is as follows. First, the sample is cooled from high temperature to the selected base temperature. Second, it is irradiated with a green laser for 20 min at an effective power of 300 μW incident on the sample (see **Supporting Information Section 2.1.3** for measurements performed with different exposure times and laser powers). Third, the laser is switched off and the system is allowed to thermally stabilize at the selected base temperature. Finally, the sample is warmed back to the desired high temperature.

**Acknowledgements**

The authors acknowledge financial support from the European Commission (ERC PoC Mol-2DMem, ref. 101287828), the Spanish Ministerio de Ciencia, Innovación y Universidades (MICIU) (Projects 2D-SPICE, ref. PID2023-149309OB-I00, cofinanced by FEDER; MICIU/AEI/10.13039/501100011033 and Excellence Unit "María de Maeztu" (ref. CEX2024-001467-M), FEDER/UE, through grant PID2023-150732NB-I00 and the Generalitat Valenciana (PROMETEO Program, CIPROM/2024/51; Ayudas para la creación de SPIN-OFFS, CISPINSEM/2026/009). SMV thanks the Spanish MICIU for the Ramón y Cajal program RYC2024-048264-I.

## Data Availability Statement

The data that support the findings of this study are openly available in Zenodo at https://doi.org/10.5281/zenodo.23168388, reference number 10.5281/zenodo.23168388.

**Supporting Information**

Supporting Information is available from the Wiley Online Library or from the author.

**Layered spin-crossover metal-organic frameworks for light-induced control of two-dimensional quantum materials**

*Carla Boix-Constant, Alejandro Orellana-Silla, José Antonio Real, Samuel Mañas-Valero,* Eugenio Coronado**

e-mail: samuel.manas@uv.es, eugenio.coronado@uv.es

ToC figure

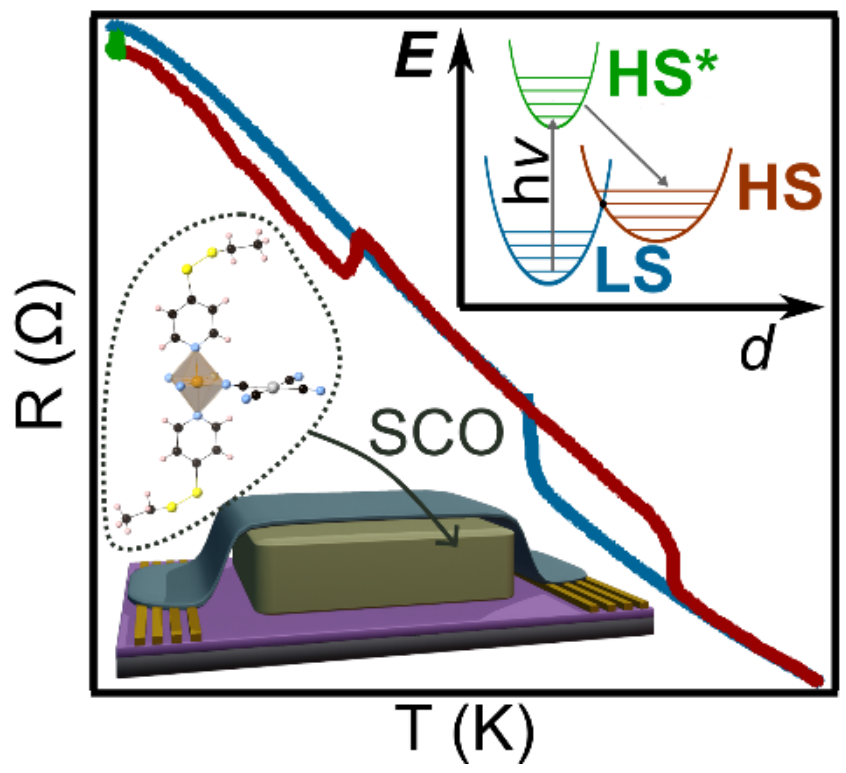

Spin-crossover metal-organic frameworks (SCO-MOFs) provide a route to control the electronic and magnetic properties of two-dimensional quantum materials. By integrating a layered SCO-MOFs with graphene, CrSBr and $NbSe_2$ nanodevices, thermally and light induced spin transitions induce strain that reversibly modulates charge transport, magnetic memory and superconducting behaviour, opening new opportunities for MOF-based electronic and spintronic heterostructures beyond conventional porous-material applications.